\documentclass{PoS}

\def\hhref#1{\href{http://arxiv.org/abs/#1}{#1}} 

\title{Connections between cosmic-ray physics, gamma-ray data analysis and Dark Matter detection}

\ShortTitle{Cosmic rays and Dark Matter}

\author{\speaker{Daniele Gaggero}\thanks{In this highlight paper I discuss
results obtained in collaboration with M.~Cirelli, G.~Di Bernardo, C.~Evoli, G.~Giesen, D.~Grasso, L.~Maccione, P.~Ullio, A.~Urbano,  M.Taoso, and M.~Valli.}\\
        SISSA, via Bonomea 265, 34136 Trieste, Italy\\
	INFN, Sezione di Trieste, via Valerio 2, 34127 Trieste, Italy\\
        E-mail: \email{daniele.gaggero@sissa.it}}

\abstract{Cosmic-ray (CR) physics has been a prolific field of research for over a century. The open problems related to CR acceleration, transport and modulation are deeply connected with the indirect searches for particle dark matter (DM). In particular, the high-quality gamma-ray data released by Fermi-LAT are under the spotlight in the scientific community because of a recent claim about a inner Galaxy anomaly: The necessity to disentangle the astrophysical emission due to CR interactions from a possible DM signal is therefore compelling and requires a deep knowledge of several non-trivial aspects regarding CR physics. I review all these connections in this contribution. 
In the first part, I present a detailed overview on recent results regarding modeling of cosmic-ray (CR) production and propagation: I focus on the necessity to go beyond the standard and simplified picture of uniform and homogeneous diffusion, showing that gamma-ray data point towards different transport regimes in different regions of the Galaxy; I sketch the impact of large-scale structure on CR observables, and -- concerning the interaction with the Heliosphere -- I mention the necessity to consider a charge-dependent modulation scenario. In the second part, all these aspects are linked to the DM problem. I analyze the claim of a inner Galaxy excess and discuss the impact of the non-trivial aspects presented in the first part on our understanding of this anomaly.
}
\FullConference{The 34th International Cosmic Ray Conference,\\
		30 July- 6 August, 2015\\
		The Hague, The Netherlands}

\begin{document}

\section{Introduction}


The apparent simplicity of the well-known equation governing cosmic-ray (CR) transport in the Galaxy captures the rich phenomenology of both the astrophysical properties characterizing the interstellar medium, and the complicated particle interactions among CRs and the ambient.
The physics standing behind the phenomenological approach is extremely complicated, and so are the environments where the process of CR propagation takes place.

Nevertheless, it has been common practice for a long time to keep the problem as simple as possible, and drop many complicated issues regarding both the process of diffusion and the description of the Galaxy: the generally accepted framework has always been the uniform and isotropic diffusion in a over-simplified model of the Galactic environment where all the main ingredients show mild variations with space and no evolution in time. This method was been well justified by the scarcity of experimental data and their poor quality: it would have been hard to constrain the parameters involved in a more complicated  description, given the poor statistics associated to CR measurements. Providing effective averages over space and time of most parameters, and keeping the picture as simple as possible has been therefore the natural option.

Nowadays, the multi-messenger approach to CR studies is getting more and more important: the study of secondary radiations such as synchrotron emission and gamma rays provides excellent datasets thanks to the effort of the experimental collaborations, and the local CR spectra are known with unprecedented precision. 
The presence of anomalies in the data and the increased accuracy changes completely the viewpoint and calls for a revision of long-standing approximations very common in the CR Literature. In general, a more realistic and accurate treatment on the modeling side is now compelling.

On top of that, the Dark Matter (DM) indirect detection problem appears strictly connected with this issue. 
The quest for signatures of particle DM annihilation or decay in CR and gamma-ray spectra has captured the attention of physicists for a long time, and some previously mentioned anomalies in the data have been interpreted as possible hints of the presence of DM itself. In particular, this has been the case of the anomalous GeV excess of gamma rays~\cite{Hunter:1997we}, not confirmed by Fermi-LAT, the positron excess at high energy identified by PAMELA~\cite{Adriani:2008zr} and confirmed by both Fermi-LAT and AMS-02, other non-confirmed anomalies in leptonic data (e.g. {\cite{Chang:2008aa}), the Galactic center excess~\cite{excess_history}
In this context, disentangling astrophysical and DM interpretation is crucial. The multi-messenger approach helps, since strong constraints from the DM scenario come from different channel. Nevertheless, in light of this problem, the already mentioned necessity for a good description of the astrophysics governing the CR propagation becomes even more important.

The purpose of this contribution is to highlight some aspects of CR propagation going beyond the standard lore, and to select some of these aspects that turn out to be relevant for DM detection.

This work is structured as follows:

\begin{itemize}

\item In the first part I discuss several aspects regarding the study of both CR spectra 
and the secondary radiation emitted by CRs themselves (namely synchrotron and gamma).
In particular, among other results, I review some relevant recent papers regarding non-trivial
properties of CR diffusion. 
%
The approach is data-driven and 
multi-messenger: I show how the evidence of non-standard diffusion regimes
stems from anomalies in the CR and $\gamma$-ray datasets. 

\item In the second part, I analyze the aspects of CR physics, among those
presented in the first section, appearing most relevant for DM indirect
detection. 
In particular, I focus on the recently claimed excess from
the Galactic Center, and show how an accurate and realistic description of CR transport
is crucial in order to scrutinize the existence of the excess itself, understand its
origin and constrain its interpretation in term of DM annihilation and decay.

\end{itemize}

\section{CRs}

\subsection{Constraining the average properties of CR transport}
\label{sec:prop_parameters}

Depending on the observables one wants to consider, and on their level
of accuracy, it may be enough to sketch a simplified picture of the Galaxy -- as
briefly described in the Introduction -- and capture the average properties of CR transport.
For example, this has been the case for many years as far as the light nuclei were concerned.
In this case, the framework described in the Introduction turned out to be satisfactory
to reproduce the spectra -- measured mainly by balloon experiments -- of protons, Helium, Boron, 
and Carbon.
In this phenomenological framework, the procedure consists in constraining the free parameters involved
in the diffusion-loss equation exploiting mainly the secondary-to-primary ratios such as
the Boron-to-Carbon (B/C) and antiproton-to-proton ($\bar{p}/p$) ratios.

This exercise was performed in the latest years by several groups.
 
{\bf 1)} In \cite{evoli:2008} the {\tt DRAGON} code was used for the first time. This package is designed to compute CR propagation in the Galaxy for all the species, both hadronic and leptonic, adopting a very general equation describing CR diffusion with spatial-dependent diffusion coefficient, and allowing for detailed three-dimensional simulations, either under the assumption of isotropic diffusion or in the most general (anisotropic) case. The code includes some ingredients taken from the public version of {\tt Galprop} such as the nuclear cross-section database, gas distribution and interstellar radiation field \cite{galprop,moskalenko:2001ya}. 
Remarkably, the results obtained in \cite{evoli:2008} were confirmed by the recently released B/C preliminary data from AMS-02 collaboration: In particular, the best-fit value for the rigidity scaling of the diffusion coefficient ($\delta=0.4$) turns out to be preferred if this dataset is considered too, as shown in \cite{evoli:2015}; however, the preliminary status of the data does not permit to make any robust statement on the uncertainty so far. 

{\bf 2)} Staying on the numerical side, in \cite{Trotta:2010mx} a detailed Bayesian analysis was performed using {\tt Galprop}~\cite{galprop,vladimirov} under simplified assumptions concerning diffusion (single power-law for the scalar diffusion coefficient with no low-energy alteration of the scaling), but --- on the other hand -- allowing breaks at arbitrary rigidity in the injection spectrum. Due to these combined assumptions, the results of this scan point towards a very narrow range for $\delta$ (around $0.33$) and select high-reacceleration models as the preferred ones. 

{\bf 3)} Concerning semi-analytical codes like {\tt USINE}~\cite{usine}, instead, several complete MCMC scans were performed (see e.g. \cite{putze:2010,maurin:2010}), and a wider class of models were shown to be compatible with the (pre-AMS) data, consistent with the results reported in \cite{evoli:2008}. In these papers a detailed study of the systematic uncertainties due, e.g., to nuclear physics, is also presented. The same group discussed the impact of AMS data in \cite{genolini:2015,giesen:2015}, showing results compatible with \cite{evoli:2015} and pointing out the role of a possible primary Boron component on the uncertainty affecting the $\delta$ parameter.

I remark that all the aforementioned works aim at fitting local CR measurements, and adopt smoothly-varying and large-scale averaged functions for most astrophysical ingredients, in particular for the CR source term~\cite{Case:1998qg,Lorimer:2003qc,FaucherGiguere:2005ny,Ferriere:2001rg}: Therefore, the models presented there must be taken {\it cum grano salis} when extrapolated to faraway regions of the Galaxy. I will come back to this issue at the end of this highlight.

\subsection{How large is the halo?}
\label{sec:halo_size}

\begin{figure}[!htb!]
\begin{center}
\centering
   \includegraphics[scale=0.37]{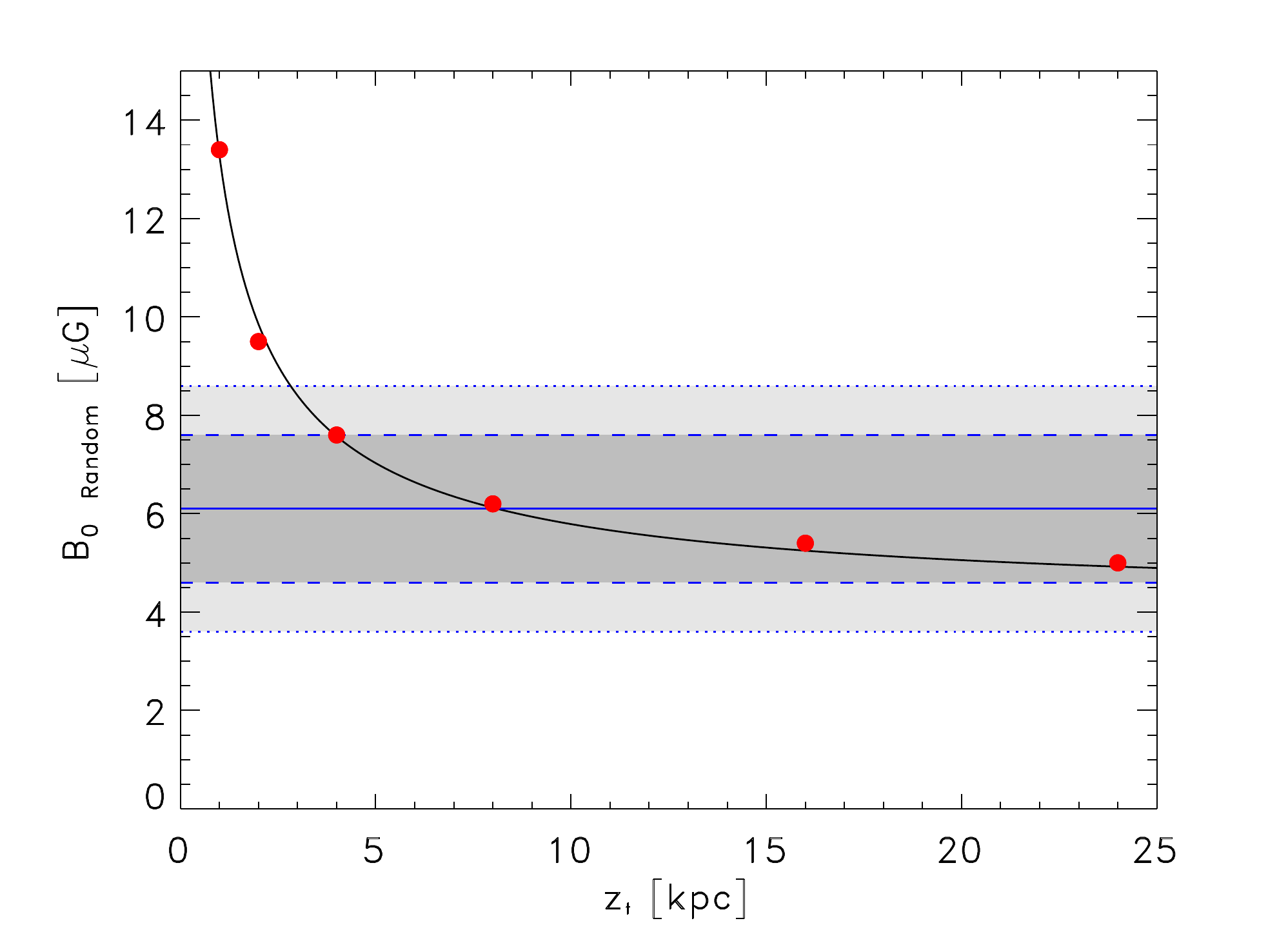}
\caption{\small \textit{The normalization of the random GMF  is plotted against its vertical scale height (which we assume to be the same as that of the diffusion coefficient).
The $3(5)~\sigma$ regions allowed by RM data are represented in gray (light-gray). Red dots are our results obtained under the condition that KRA models reproduce the observed synchrotron spectrum.  The black line is a fit of those points. 
Plot taken from \cite{dibernardo:2013}.}}
\label{fig:halo_size}
\end{center}
\end{figure}

The extension of the diffuse halo in the vertical direction is still unknown; since this parameter plays a crucial role for the predictions regarding DM models, I will present some details.
The halo size $L$ can be constrained looking at radioactive isotopes: In particular, the ratio $^{10}{\rm Be}/^{11}{\rm Be}$ (see e.g. the review paper \cite{strong:2007nh} and references therein). 

However, due to the poor quality of current data, and to the huge uncertainty associated to solar modulation, a wide range of values is allowed: A multi-messenger approach is therefore compelling. In particular, the synchrotron data provide interesting information for this purpose. In \cite{dibernardo:2013} we analyzed both mid-latitude spectra and profiles. The profiles clearly show a preference for thick halos, but the errors are too large to get strong constraints. On the other hand, if one assumes an exponential decay of the turbulent magnetic field with $z$, and taking into account the precise local measurements of this observable with associated uncertainty, it was possible to show that the normalization of synchrotron spectra from $\sim 10$ to $\sim 10^5$ MHz, integrated in the region $10^{\circ}~<~b~<~20^{\circ}$, provide a quite stringent constraint on $L$, as shown in Fig. \ref{fig:halo_size}. The preference for thick halos is confirmed, and values smaller than $2$ kpc are excluded under these assumptions. 

We will discuss in sec.~\ref{sec:constraints} the implication of this results on DM constraints.

\subsection{Structures in the Galaxy and their impact on CR propagation}

\begin{figure}[!htb!]
\begin{center}
\centering
   \includegraphics[scale=0.17]{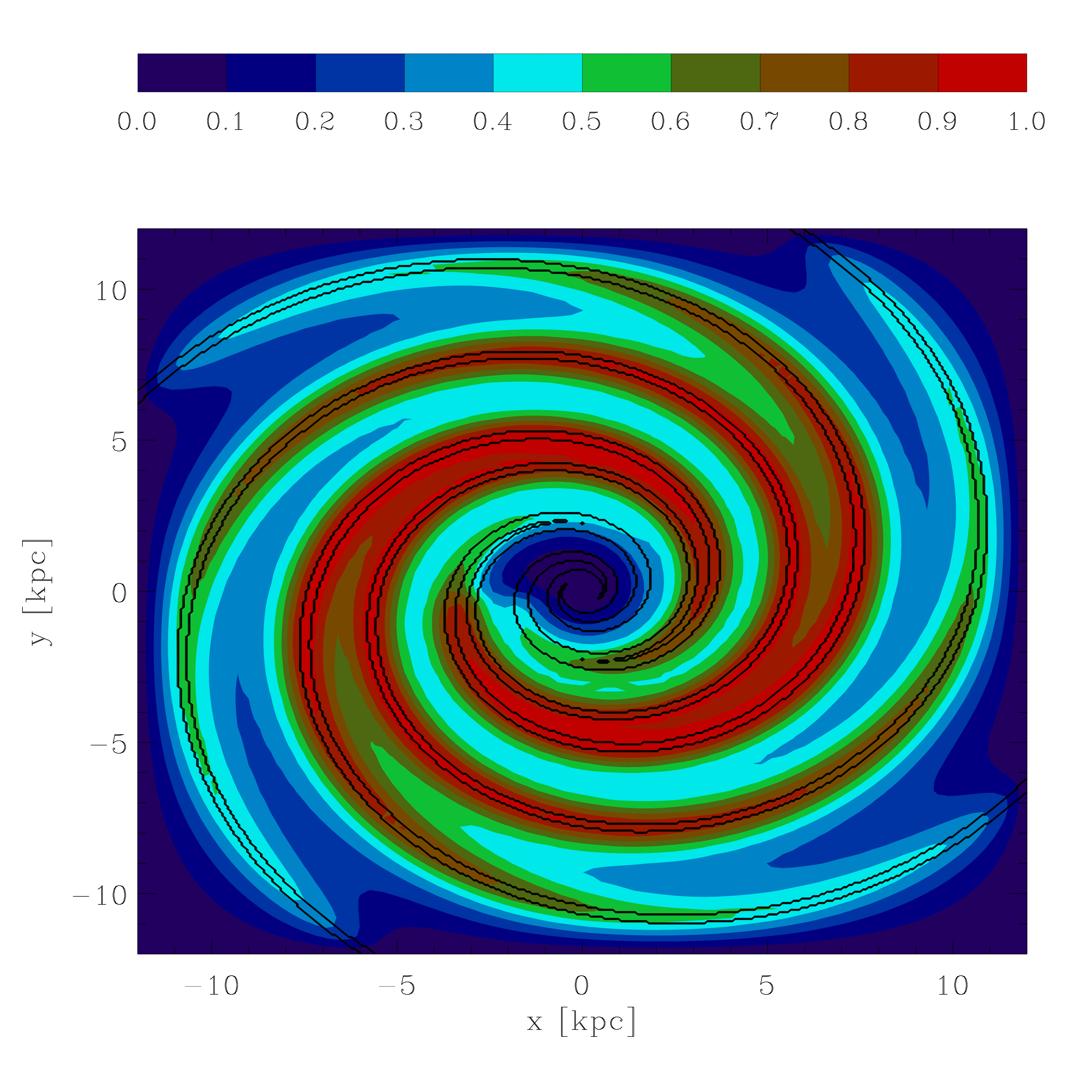}
\caption{\small \textit{Face-on view of the propagated 100 GeV electron distribution on the Galactic plane for
sources distributed in the spiral arms.}}
\label{fig:3D}
\end{center}
\end{figure}

All the results described so far are based on a simplified description of the Galaxy as a cylinder with no structure at all: All
the relevant quantities are smooth functions of $R$ and $z$. Instead, it is well known that the Galaxy has a spiral structure.  There has been a general consensus on the existence of this pattern between $\sim 3$ and $\sim 10$ kpc, although some disagreement remains on the details regarding the number of spiral arms and their geometry, traced either by catalogs of point sources~\cite{Wainscoat:1992pk} or by the study of the far-infrared cooling lines of the interstellar gas~\cite{SteimanCameron:2010tm}.

This pattern cannot be safely neglected when modeling the propagation of high-energy leptons. Because of the severe energy losses (mainly due to synchrotron emission and Inverse Compton scattering), electrons and positrons above $100$ GeV are expected to come from quite nearby regions (the horizon is less $\sim 1$ kpc at $\sim 1$ TeV), and therefore their flux is likely to be strongly influenced by the aforementioned three-dimensional large-scale structure. Driven by these considerations, in~\cite{gaggero:2013} we were able to reproduce all the local CR spectra measured by AMS-02 adopting a source function featuring the spiral pattern taken from \cite{Wainscoat:1992pk}. Due to our position in an interarm region, the energy losses are enhanced in this scenario with respect to the smooth source term case, because the electrons have to travel longer distance to reach us if the sources are located in the arms: Therefore, it is possible to reproduce the leptonic data with a harder injection index, in less strong tension with that predicted from the theory of Fermi acceleration. We show in Fig.~\ref{fig:3D} the face-on map of the propagated high-energy electrons for illustrative purposes.
The impact of the spiral structure on CR observables, with focus on the secondary-to-primary ratios and anisotropy, is also discussed in detail in ~\cite{Kissmann:2015kaa}, exploiting the recently introduced {\tt PICARD} code; the impact on the B/C found in that paper appears compatible with our findings.

\subsection{Data-driven hints of non-uniform CR transport}
\label{sec:diffusion}


Diffusion is the most important process governing CR transport.

As mentioned in the Introduction, all the results in the literature, and all those discussed so far in this highlight, have been obtained under the hypothesis of constant and homogeneous diffusion.
This approximation is clearly too rough, given the extreme variability of the ISM properties from one region to the other. 
Nevertheless, it has turned out to be satisfactory in particular for the treatment of light nuclei, since they are mostly
sensitive to the average properties of the ISM.
The situation changes dramatically when one takes into account the gamma-ray emission: Since this radiation traces the CR distribution through the Galaxy, non-local properties become crucial, and in principle it is possible to trace variations of CR transport regime by analyzing this radiation.

\subsubsection{Gradient Problem} In \cite{evoli:2012} a first attempt to deal with these issue is presented. The motivation
of this analysis is the presence of two long-standing anomalies: The {\it gradient problem} -- i.e. the well-known fact that
the CR radial gradient in the vicinity of the Sun appears much flatter than predicted \cite{gradient1,gradient2,gradient3} -- and the {\it anisotropy problem}, i.e. the larger dipole anisotropy predicted by numerical models (especially with $\delta > 0.5 \div 0.6$) compared to the measured one (on this topic, see the detailed discussion in \cite{Blasi:2011fm}).

The physical idea we consider stems from the different behavior of the diffusion coefficients in the parallel and perpendicular direction with respect to the regular magnetic field: the former is expected to decrease with increased turbulence, while the latter is expected to decrease, as predicted by quasi-linear theory and confirmed by numerical simulations~\cite{DeMarco:2007ux}. Bearing this information in mind, and taking into account that: {\it 1)} the perpendicular escape is dominant (at least for $R<R_{\rm Sun}$), {\it 2)} larger turbulence is expected where more CR sources are present, we proposed a two-dimensional phenomenological model where the diffusion coefficient $D(R)$ is correlated to the CR source term $Q(R)$ via the general expression $D \propto Q^{\tau}$. 

Remarkably, both problems are solved in this framework, while the local CR observables are consistently reproduced. The flattening of the gradient predicted from our model can be seen in Fig.~\ref{fig:gradient}, compared to the gradient inferred from Fermi-LAT observations.

\begin{figure}[!htb!]
\begin{center}
\centering
   \includegraphics[scale=0.35]{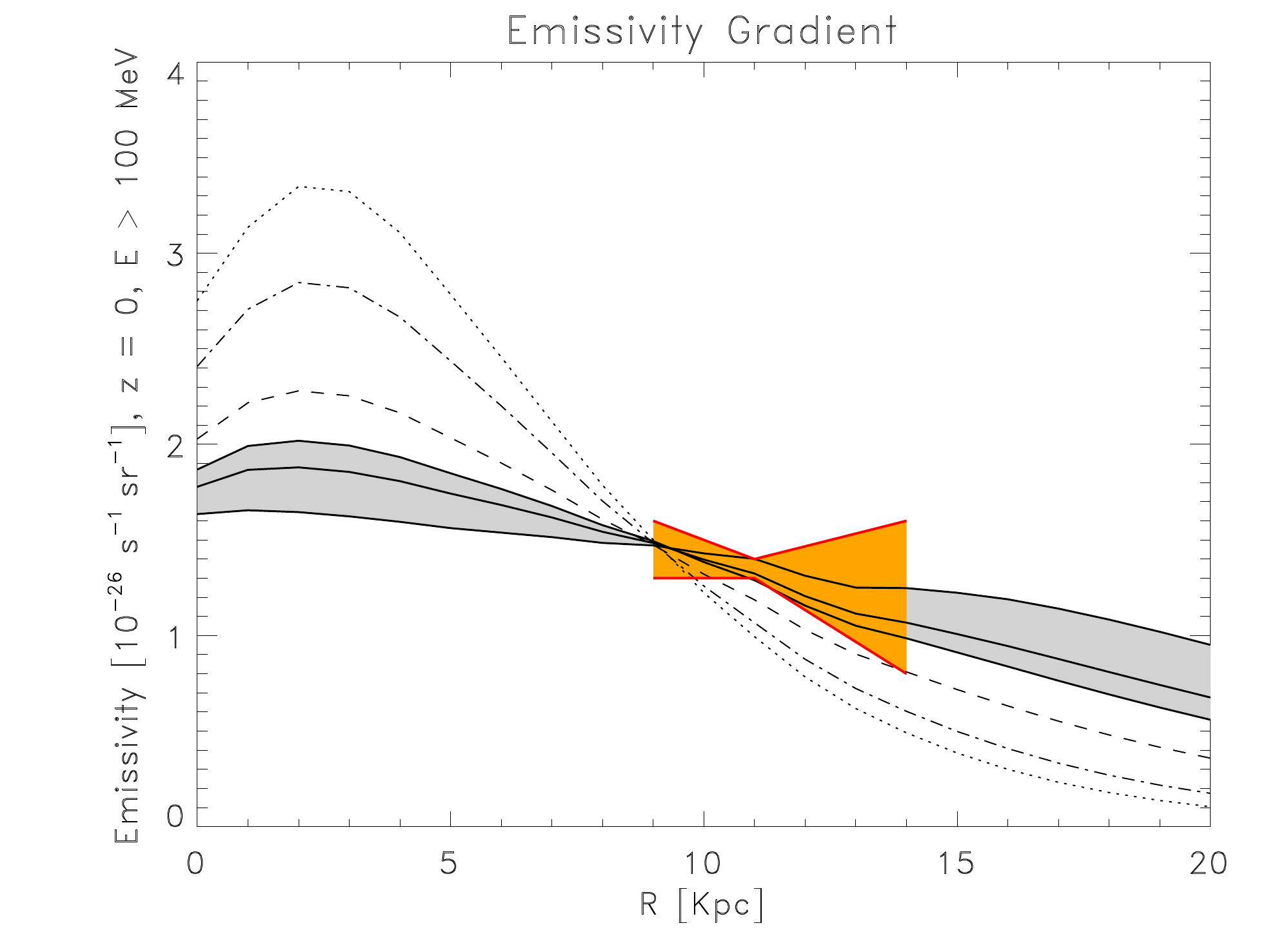}
\caption{\small \textit{Integrated $\gamma$-ray emissivity (number of photons emitted per gas atom per unit time) constrained by Fermi-LAT (orange region \cite{gradient2}, grey region \cite{gradient3}) compared with our predictions for  $\tau=0,0.2,0.5,0.7,0.8,0.9$ (from top to bottom).}}
\label{fig:gradient}
\end{center}
\end{figure}

\subsubsection{Slope Problem} Another intriguing hint of a non-trivial behavior of the diffusion coefficient comes from the discrepancy -- mentioned in \cite{FermiPaper} -- between the high-energy gamma-ray slope predicted by conventional numerical simulations along the Galactocentric plane and the measured one. This tension becomes larger with decreasing longitude, and disappears at large latitude. In \cite{gaggero:2015} we interpreted this in terms of an increasing $\delta$ with $R$ (see also \cite{Erlykin2012}). For the first time, starting from this hypothesis, we were able to produce a self-consistent propagation model that correctly reproduces both the gamma-ray spectra in all the relevant sky windows along the Galactic plane and the local CR measurements (see fig.~\ref{fig:KRAgamma} for the inner Galaxy). This behavior may be interpreted as the signature of a smooth transition between a parallel escape along the vertical component of the magnetic field in the inner Galaxy (see \cite{farrar}) and a regime of perpendicular escape at larger radii, governed by the (steeper) perpendicular diffusion coefficient. I point out that -- as shown in \cite{Gaggero:2015xza,Gaggero:2015nga,Gaggero:2015jma} -- it is possible to test this scenario with the forthcoming neutrino experiments, and it is already possible to notice that, under this framework, the long-standing TeV anomaly detected by Milagro can be explained consistently.

\begin{figure}[!htb!]
\begin{center}
  \includegraphics[scale=0.35]{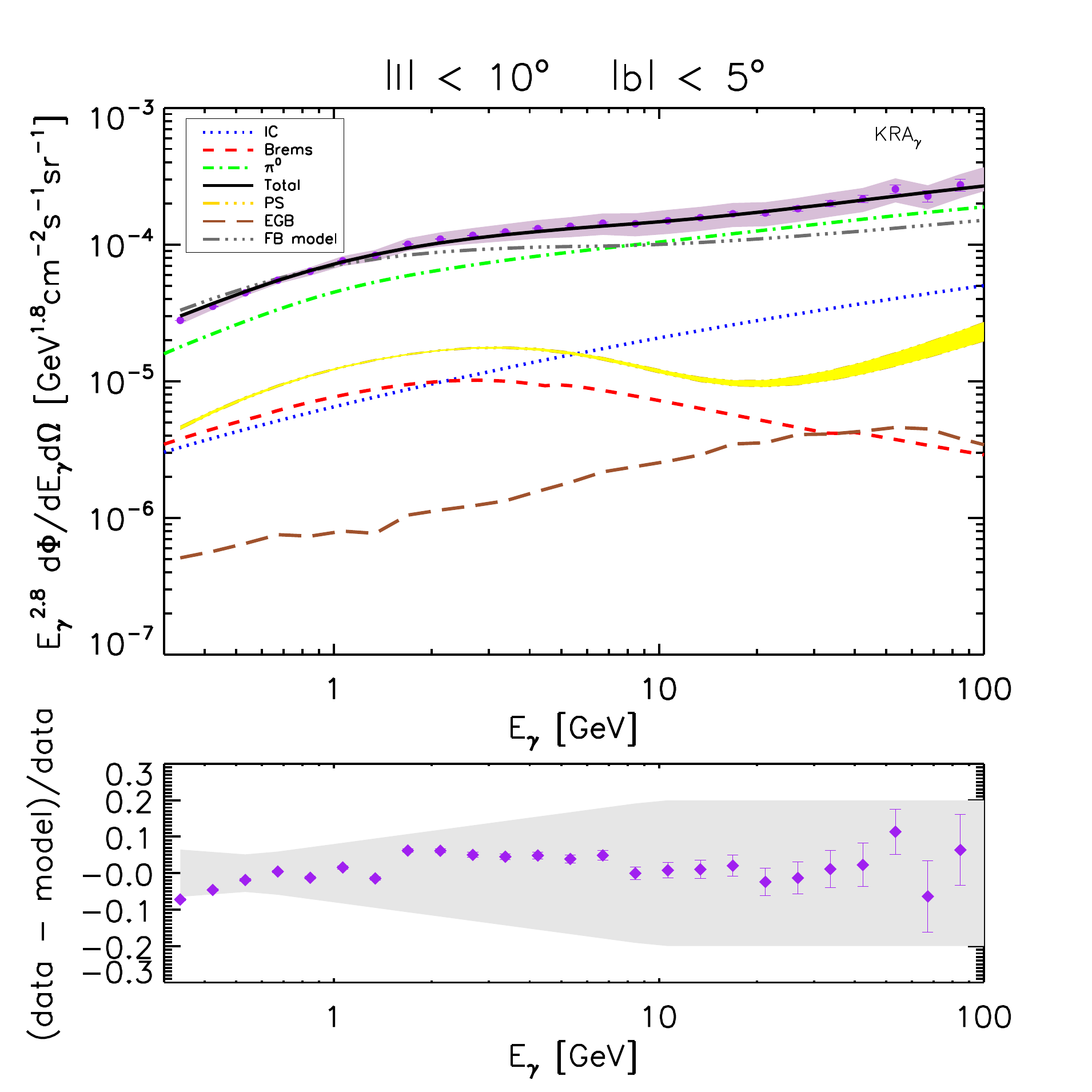}
\caption{\small \textit{ Comparison between the $\gamma$-ray flux computed with the CR propagation model proposed in \cite{gaggero:2015} (a modified Kraichnan model labeled KRA$_{\gamma}$). The solid black line is the corresponding total flux; individual components shown) and the {\tt Fermi-LAT} data (purple dots, including statistic and systematic errors)
in the Galactic disk. For comparison, we show the total flux for the Fermi Benchmark (FB) model defined in ref.~\cite{FermiPaper} (double dot-dashed gray line). Lower panel. Residuals computed for KRA$_{\gamma}$ and FB models.}}
\label{fig:KRAgamma}
\end{center}
\end{figure}

\subsection{The final stage of the voyage: The Heliosphere}
\label{sec:modulation}

When charged CRs penetrate into the Heliosphere, they are subject to the solar modulation effect, very relevant
for low-energy particles (below $\simeq 10$ GeV). The complicated between CRs and the solar wind -- a
complicated combination of diffusion, convection, magnetic drift and adiabatic energy losses -- is usually 
parametrized by the well-known {\it force-field} formula \cite{Gleeson:1968zza}.
However, a more detailed and realistic treatment, based on a set of stochastic differential equations, 
can be found in \cite{potgieter}. In \cite{helioprop} it was shown that -- given the accuracy of the current leptonic data --
the different low-energy positron datasets corresponding to different data-taking periods can be successfully reproduced in
a consistent way with {\tt HelioProp}, a numerical code based on the aforementioned formalism. It is now well accepted that, in order to 
reproduce correctly PAMELA, AMS-02 and Fermi-LAT low-energy datasets it is compelling to go beyond the simplified
force-field approximation.

\section{Dark matter indirect detection}

\subsection{Connecting DM search and CRs: General considerations}

The fascinating debate going on within both the astrophysics and particle physics communities about the nature
of the massive halos providing support to all the observed galaxies brought many scientists to turn their attention
to cosmic rays and gamma rays.
In particular, the most beaten track of the weakly-interacting massive particle (WIMP) hypothesis is particularly appealing from this point of view. Given the alleged properties and mass range of this class of candidates, it is natural to expect some yield of standard-model particles
produced by their self annihilation in the inner part of the Galaxy. 
These considerations led to consider the so-called indirect dark matter search as a promising field of research.

\subsection{The claim of a inner Galaxy excess}

The first papers claiming an anomalous gamma-ray emission coming from a region around the Galactic center and extending up to
$\simeq 10^\circ$ in latitude and longitude date back to $2009$~\cite{excess_history}. 
The presence of the excess was confirmed, and
its features were robustly characterized, by a comprehensive scan~\cite{Calore} based on {\tt Galprop} models for the $\gamma$-ray 
diffuse diffuse emission originating from the standard interactions of hadronic and leptonic CRs with the interstellar environment.
The existence of the anomaly does not come from a direct comparison between a given model and data, but relies on the so-called {\it template-fitting algorithm}: For each energy bin, the spatial templates associated to the different gamma-ray emission mechanisms are independently rescaled in order to obtain the best fit of the data. The exercise is then repeated with a DM-like spherical symmetrical template, and the purpose is to check whether it is used by the algorithm. Performing this method for each energy, one can derive a spectrum associated to the extra template.
The spectral shape obtained this extra emission points towards a natural interpretation in terms of DM annihilation. Intriguingly, the WIMP annihilation cross section needed to reproduce the anomaly is close to the one expected within the well-established framework of
thermal decoupling in the early Universe.

However, many alternative astrophysical interpretations were pointed out~\cite{MSP,Petrovic:2014uda,Carlson:2014cwa,YusefZadeh:2012nh,Macias:2013vya,Cholis:2015dea}. In particular, according to the independent analyses presented in \cite{Bartels:2015aea} (based on a wavelet decomposition of the gamma-ray sky) and \cite{Lee:2015fea} (based on a generalized template fit accounting for non-Poissonian photon statistics), a population of unresolved point sources seems preferred with respect to the DM scenario, and millisecond pulsars would be good candidates for this class of sources. 

\subsection{The necessity of a better description of the GC environment: sources, diffusion}

\begin{figure}[!htb!]
\centering
\minipage{0.48\textwidth}
  \includegraphics[width=1.\linewidth]{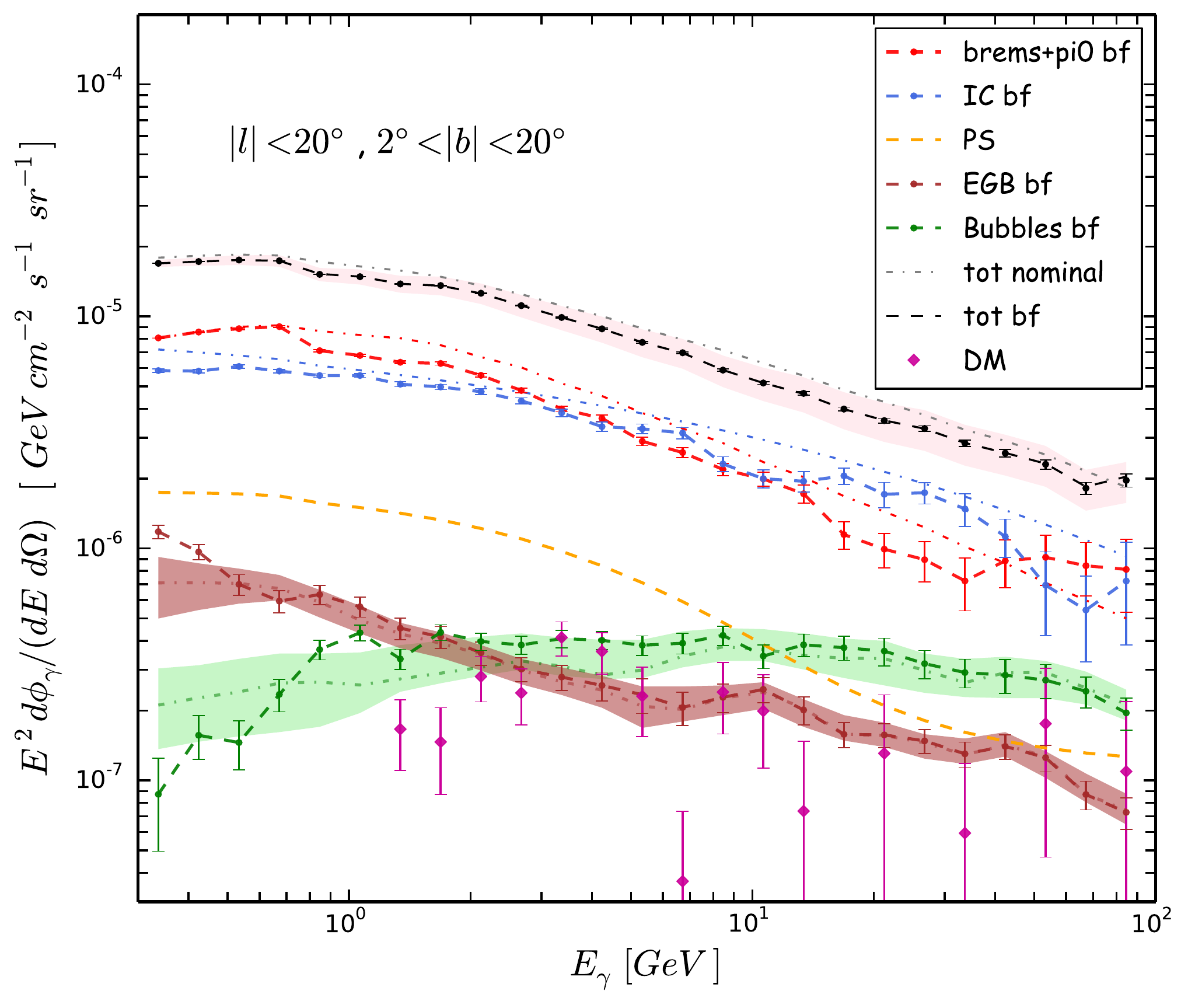}
\endminipage\hfill
\minipage{0.48\textwidth}
  \includegraphics[width=1.\linewidth]{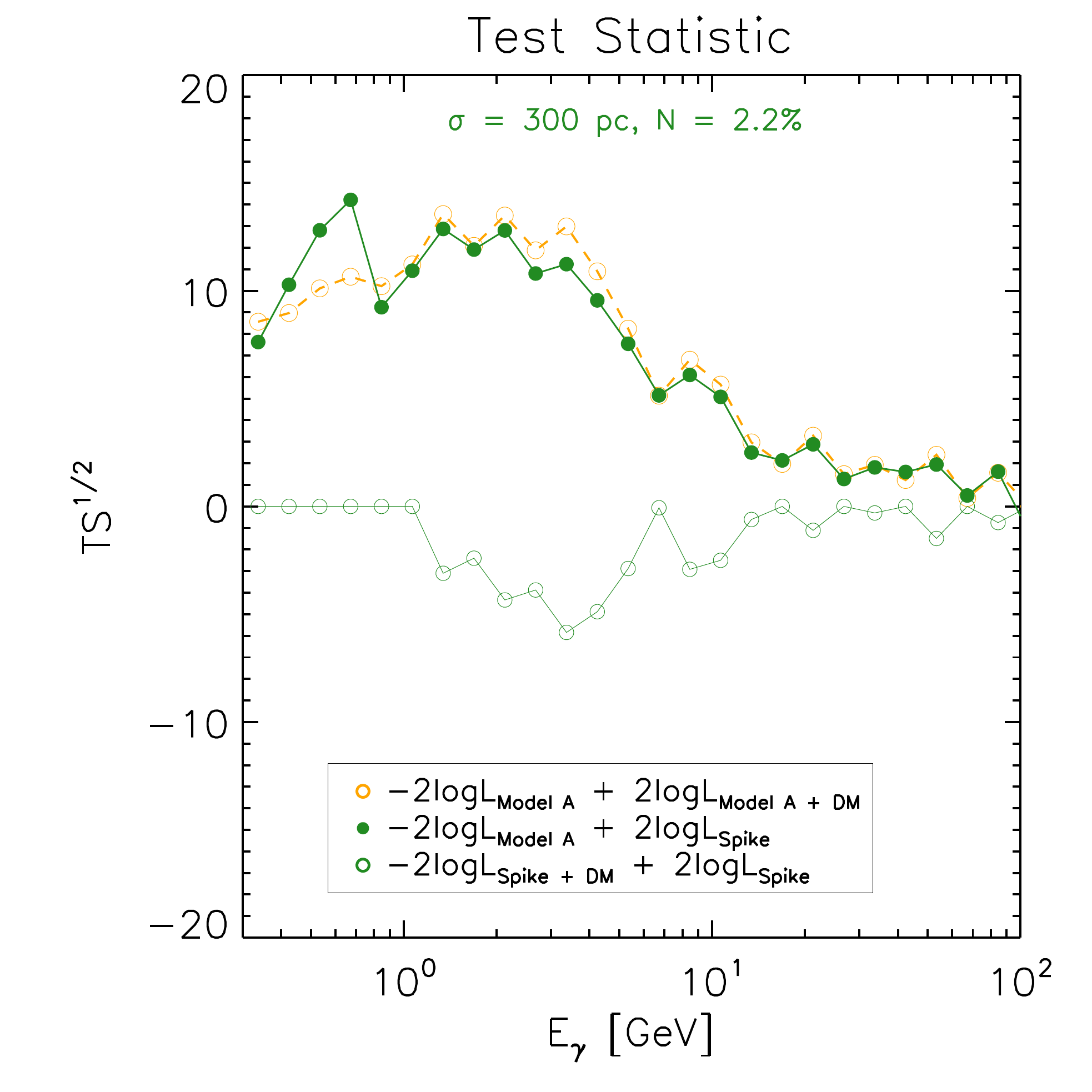}
\endminipage
\caption{\small 
\textit{{
\bf Left panel:} Spectrum of the various contributions to the total $\gamma$-ray flux, pre- and post-template-fitting, compared to the Fermi-LAT data in the ROI characterized by $|l| < 20^{\circ}$, and $2^{\circ} < |b| < 20^{\circ}$. The violet band superimposed to the data represents the systematic uncertainty. For Inverse Compton (light blue), $\pi^0$+Bremsstrahlung (red), isotropic background emission (dark red) and Fermi bubbles (green) dot-dashed lines show the nominal spectrum (pre-fitting) while points and dashed lines are the post-fitting values. Taken from~\cite{Gaggero:2015nsa}.
{\bf Right panel:} We compare the test-statistic (${\rm TS} = -2\Delta\log\mathcal{L},$  we show the square-root of TS) of the models we consider; a positive difference between two models means that the second model performs better. Yellow filled circles: $-2\log\mathcal{L}_{\rm Model\,A}+2\log\mathcal{L}_{\rm Model\,A + DM}$. Green filled circles: $-2\log\mathcal{L}_{\rm Model\,A}+2\log\mathcal{L}_{\rm Spike}$. Green empty circles: $-2\log\mathcal{L}_{\rm Spike + DM}+2\log\mathcal{L}_{\rm Spike}$. Taken from~\cite{Gaggero:2015nsa}
}}
\label{fig:likelihood}
\end{figure}

\begin{figure}[!htb]
\minipage{0.33\textwidth}
  \includegraphics[width=1.\linewidth]{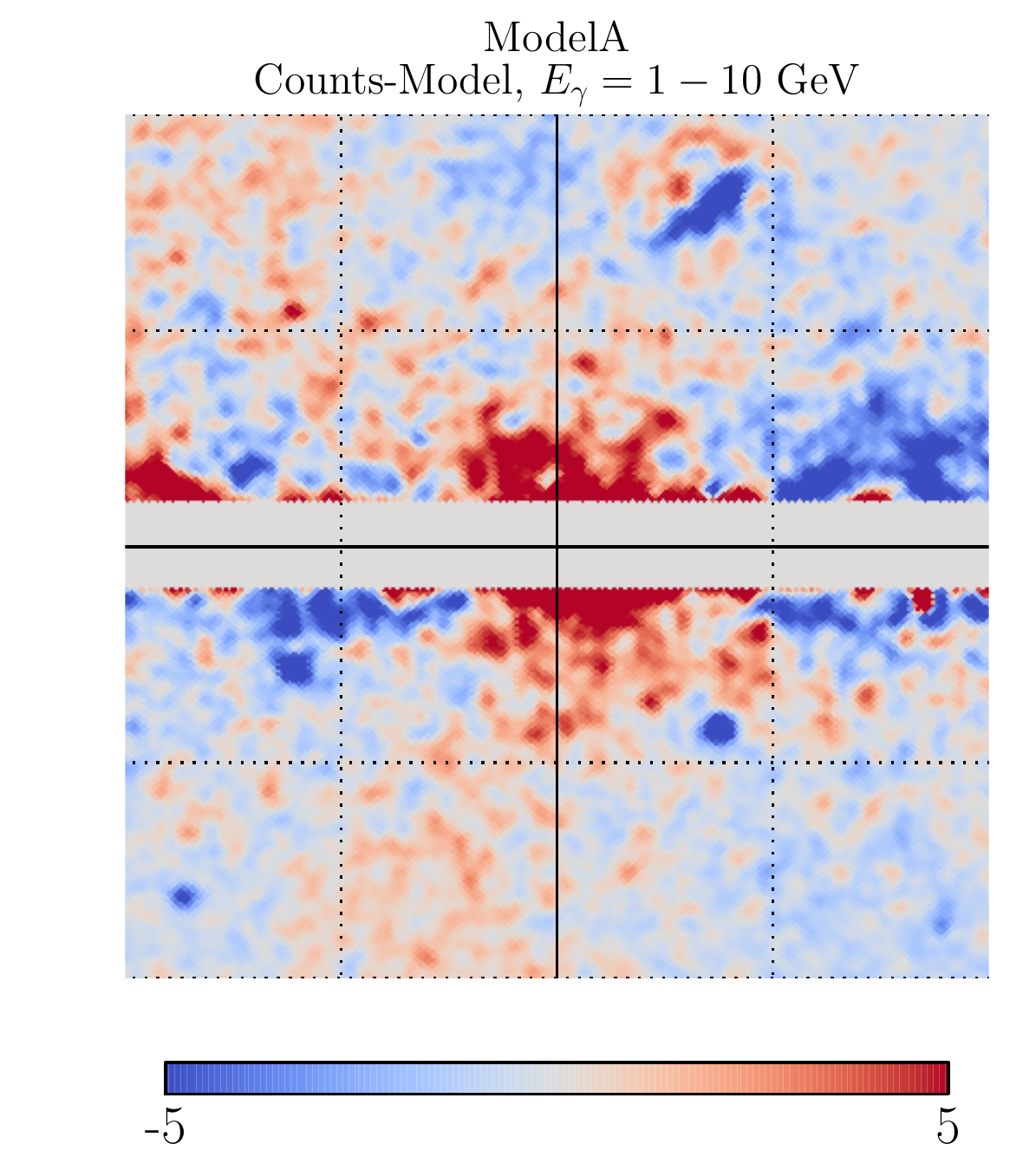}
\endminipage\hfill
\minipage{0.33\textwidth}
  \includegraphics[width=1.\linewidth]{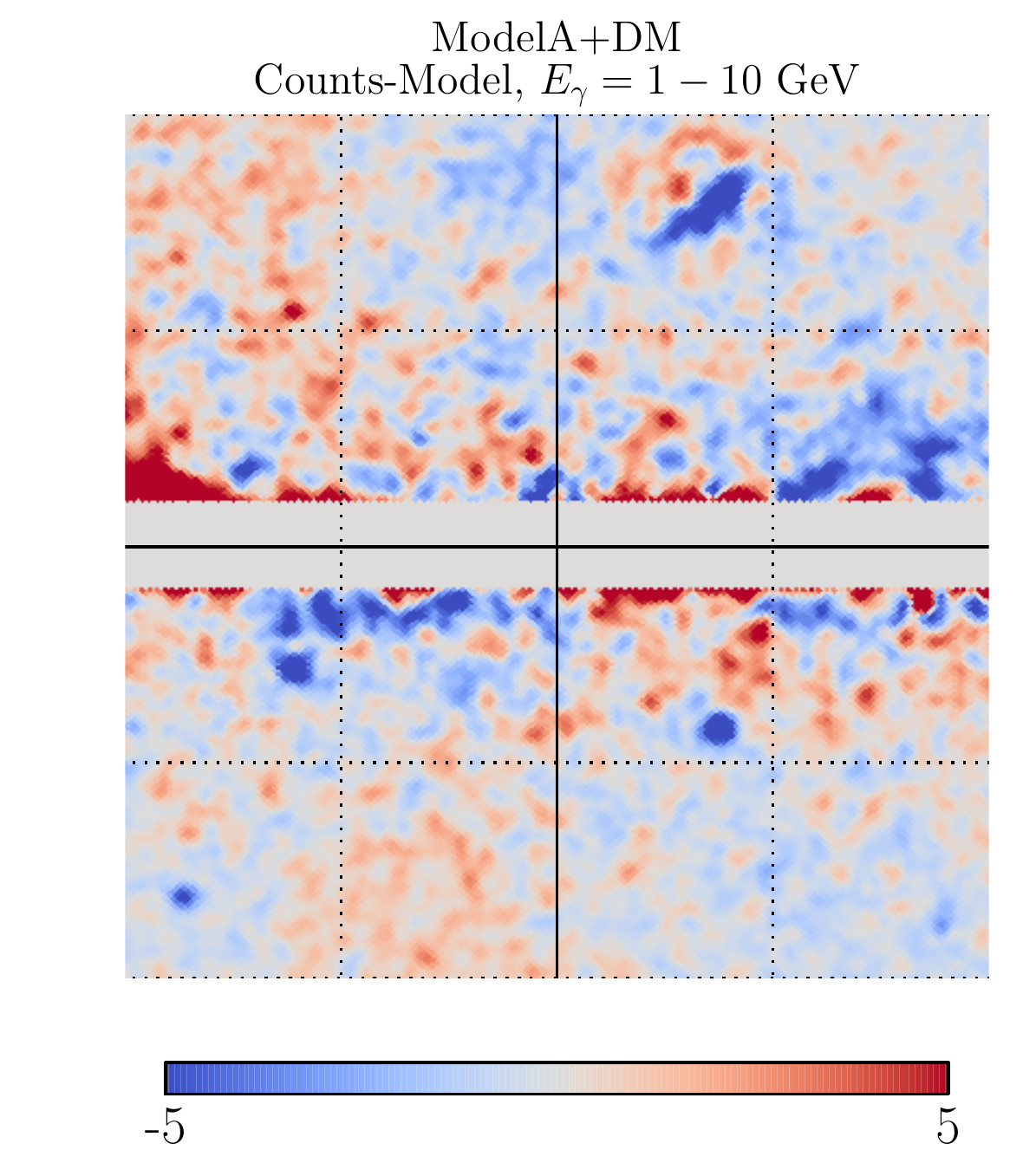}
\endminipage\hfill
\minipage{0.33\textwidth}
  \includegraphics[width=1.\linewidth]{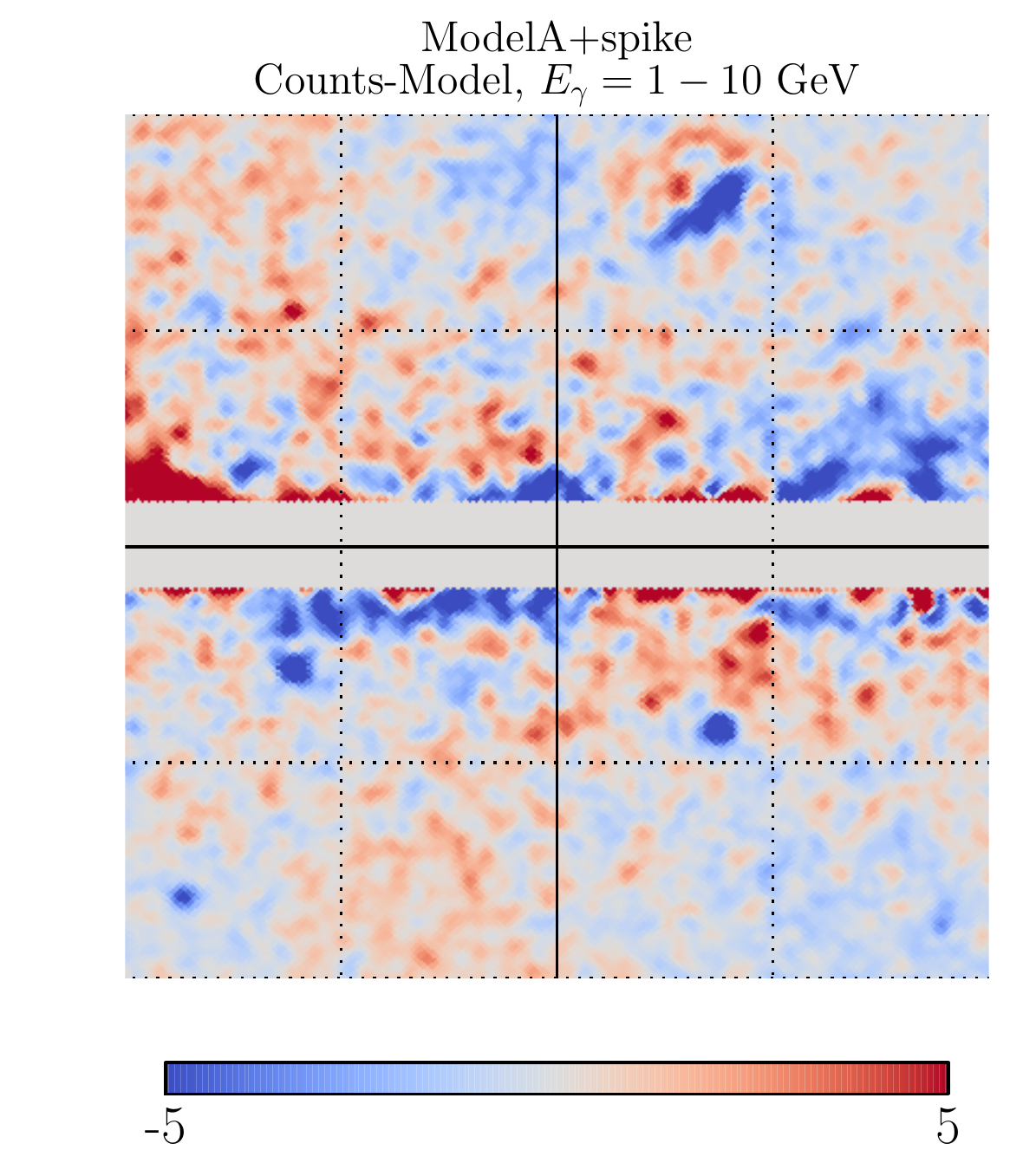}
\endminipage
 \caption{\small \textit{
Residual counts obtained for the Model A (left panel,  without the inclusion of DM template), 
for the Model A + DM (central panel), and for the spike case (right panel, without the inclusion of DM template). See text for a detailed discussion. Taken from~\cite{Gaggero:2015nsa}.
  }}
 \label{fig:residuals}
\end{figure}

The point we want to make here originates from the most natural question regarding this problem: Are we modeling the astrophysical emission correctly?

The answer is most likely negative. Indeed, the standard CR production-propagation framework considered so far, as sketched in sec. \ref{sec:prop_parameters}, is not optimized for the inner Galaxy region, and its main purpose is to provide a good fit to the local observables.
The main ingredient under scrutiny should be the CR source term. The fitting functions extrapolated from catalogs of SNRs, pulsars, or other tracers of ongoing star formation, mostly drop to zero (or to low values) in the GC. This is in strong disagreement with the evidence -- coming from infrared and X-ray observations -- of a significant star formation going on in an extended central region~\cite{Figer:2002bi,star_formation_GC,Figer:2008kf} correlated with a large reservoir of molecular gas extending up to $\sim 200\div300$ pc away from the GC~\cite{central_molecular_zone}.
These observations suggest to reconsider to the usual ingredients adopted in standard CR propagation setups, and revise them in order to make the description of the central more detailed.
Moreover, all the arguments discussed in sec.~\ref{sec:diffusion} play a major role in this context. In particular, since the inner bulge seems to host strong magnetic fields ($\sim 50 \div 100$~$\mu$G~\cite{crockernature}) compared to the estimated average value within the disc or the halo, both the structure and the effective values of the diffusion tensor are expected to show non-negligible departure from isotropy, homogeneity and from the standard values adopted in the literature. 
The same argument holds for advective effects, whose anomalous strength is observed in this region~\cite{crocker2011a,crocker2011b}.

Some considerations of this kind were raised in \cite{Carlson:2014cwa}. 
In this work, the authors mention the possibility of a point-like CR source at the GC location, possibly connected with the activity of the supermassive black hole (associated to the Sgr A$^\star$ radio source). 
Both the steady-state and the bursting regime were considered, and the authors concluded that a series of bursting events is favored with respect to a continuously emitting central point source. This motivated further investigations \cite{Petrovic:2014uda,Cholis:2015dea} which led to the conclusion that the quality of the fit in a single or multiple burst scenario is worse than the DM case.

In our recent work \cite{Gaggero:2015nsa}, instead, we consider a modified source term featuring a more extended central CR source (modeled as a Gaussian), with a size comparable to the Central Molecular Zone (CMZ). With this modification of the standard scenario, we reconsidered the template-fitting algorithm. The idea was to start with a propagation setup ({\it Model A} from~\cite{Calore}) in which: {\it 1)} The template-fitting procedure does not alter the spectra in a significant way; {\it 2)} After the fitting procedure is performed {\it without} the DM-like template, a spherical-symmetric residual is clearly seen in the comparison between fitted model and data. Once we put ourselves in this ``worst-case scenario'', we tested whether our alternative picture could perform equally well. 
Looking at Fig.~\ref{fig:likelihood} (left panel) one can see that, with this modification of the source term, the contribution of the DM template is significantly reduced, being consistent with zero (within the error bars) in  a large energy range, and -- most importantly -- gives rise to a featureless spectrum.
The main results can be seen in Fig.~\ref{fig:likelihood} (right panel): Our model, in spite of the very simplified modelization of the source term, performs as well as the DM case from the point of view of the goodness of the fit. In Fig.~\ref{fig:residuals}, instead, it is possible to appreciate the presence of an excess distributed around the GC (left panel), which is reabsorbed both in the DM case (central panel) and in our scenario (left panel).

Although some details still need to be fully understood, the remarkable point is that the energetic required to sustain the gamma-ray flux  needed to reabsorb the excess is compatible with the (independent) order-of-magnitude estimate of the ratio between the star formation rate in the central region of the Galaxy (within few hundred parsecs) compared to the total one in the Galaxy. Moreover, the results appear solid against several variations of the scenario, and hold for different sizes of the central sources, down to $\simeq 200$ and up to $\simeq 400$ pc. We did not improve significantly the fit when implementing anisotropic diffusion, but -- bearing in mind all the previously discussed arguments -- a further investigation in this direction is compelling in our opinion. 

It is clear from these considerations and results that all the caveats presented in the first part of this highlight turn out to be very important when one tries to investigate how robust an indirect DM claim is: For the specific case of the excess from the inner Galaxy, the energy budget and spatial distribution of all the different kinds of CR acceleration sites in the inner bulge play a crucial role. Both the possibilities of a major contribution from standard SNRs correlated to the CMZ and from an unresolved population of millisecond pulsars are appealing, and a combination of the two scenarios is also viable. 
Moreover, in order to shed light on this topic, a more detailed assessment of the diffusion properties, although very challenging, appears unavoidable. 	

\subsection{Constraining the DM interpretation: the role of the astrophysical uncertainties: CR transport, halo size, modulation}
\label{sec:constraints}

\begin{figure}[!htb!]
\begin{center}
\centering
\minipage{0.49\textwidth}
   \includegraphics[width=1.\linewidth]{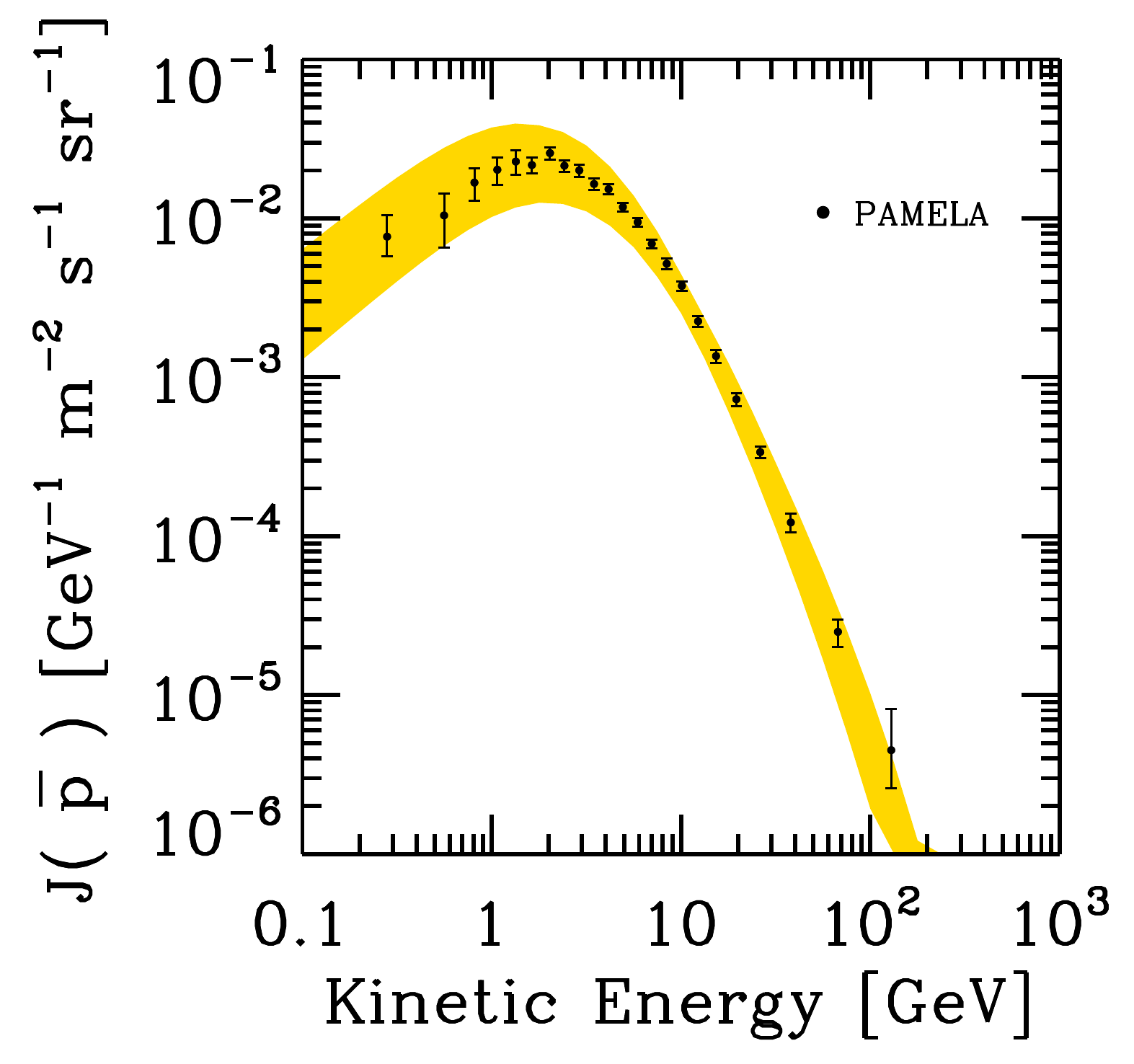}
\endminipage\hfill
\minipage{0.49\textwidth}
   \includegraphics[width=1.\linewidth]{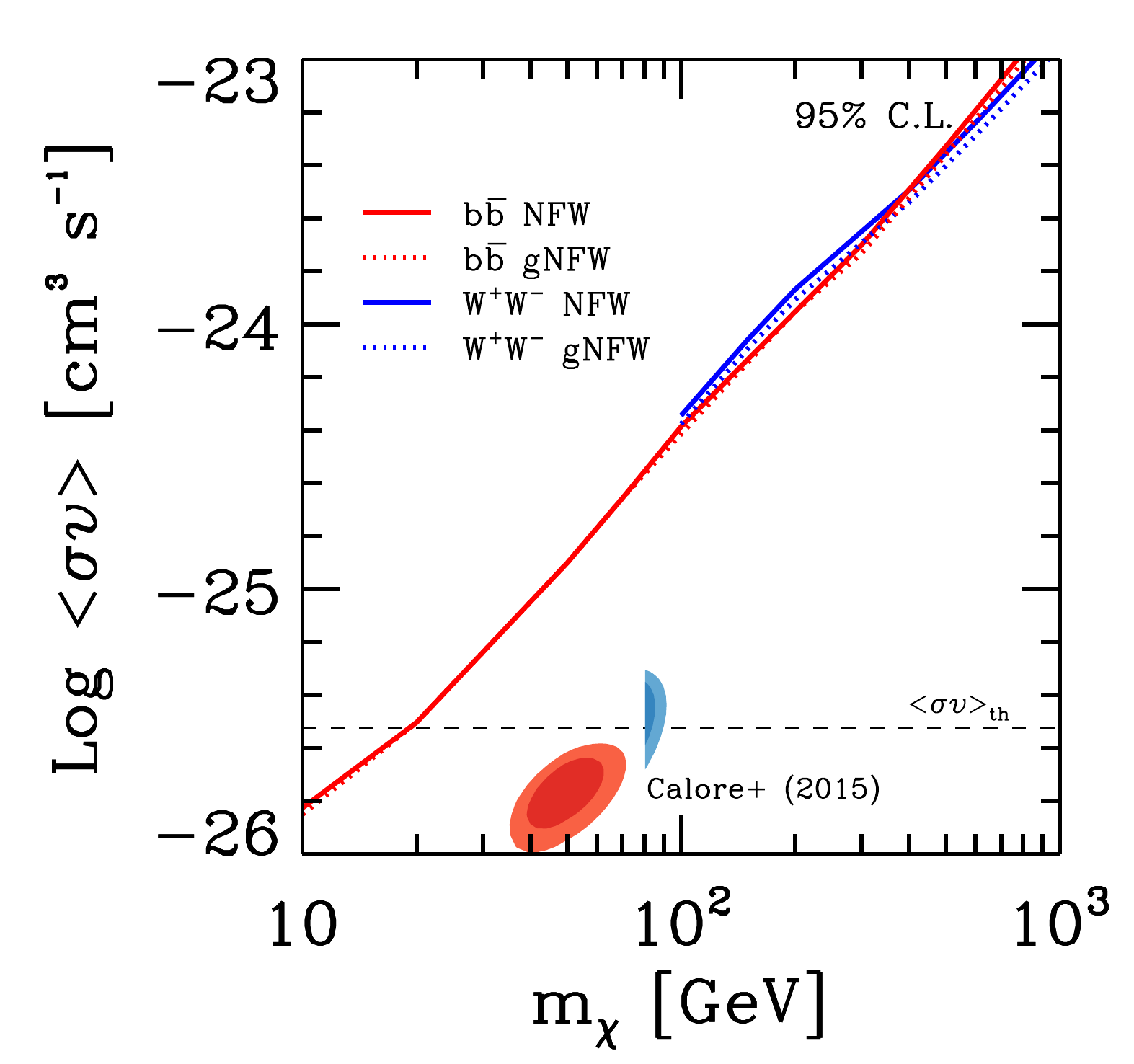}
\endminipage
\caption{\small \textit{{\bf Left panel:} The envelope of the secondary antiproton spectra computed with the different propagation
models found to reproduce the B/C and primary spectra. {\bf Right panel:} Antiproton bounds on DM models computed with conservative assumptions for the CR propagation and DM contribution. {\it Red solid (dashed) line:} $b \bar{b}$ channel for NFW (generalized NFW) profile; {\it blue solid (dashed) line:} $W^+ W^-$ channel for NFW (generalized NFW) profile. With red (blue) filled contours we report the $2\sigma$ and $3\sigma$ best-fit regions identified in~\cite{Calore} for the $b \bar{b}$ ($W^+ W^-$) channel.  Plots taken from \cite{evoli:2015}}}
\label{fig:pbar_uncertainty}
\end{center}
\end{figure}

The antiproton channel plays a crucial role as far as DM constraints are concerned, both for hadronic and -- due to
weak bremsstrahlung -- for leptonic channels (but in the latter case only for very large DM mass, $\sim 1 \div 10$ TeV). 
This is due to the astrophysical background quite under control: the only relevant component comes as a secondary yield of the 
collisions between CR nuclei and interstellar gas.

\begin{figure}[!htb!]
\minipage{0.32\textwidth}
  \includegraphics[width=1.\linewidth]{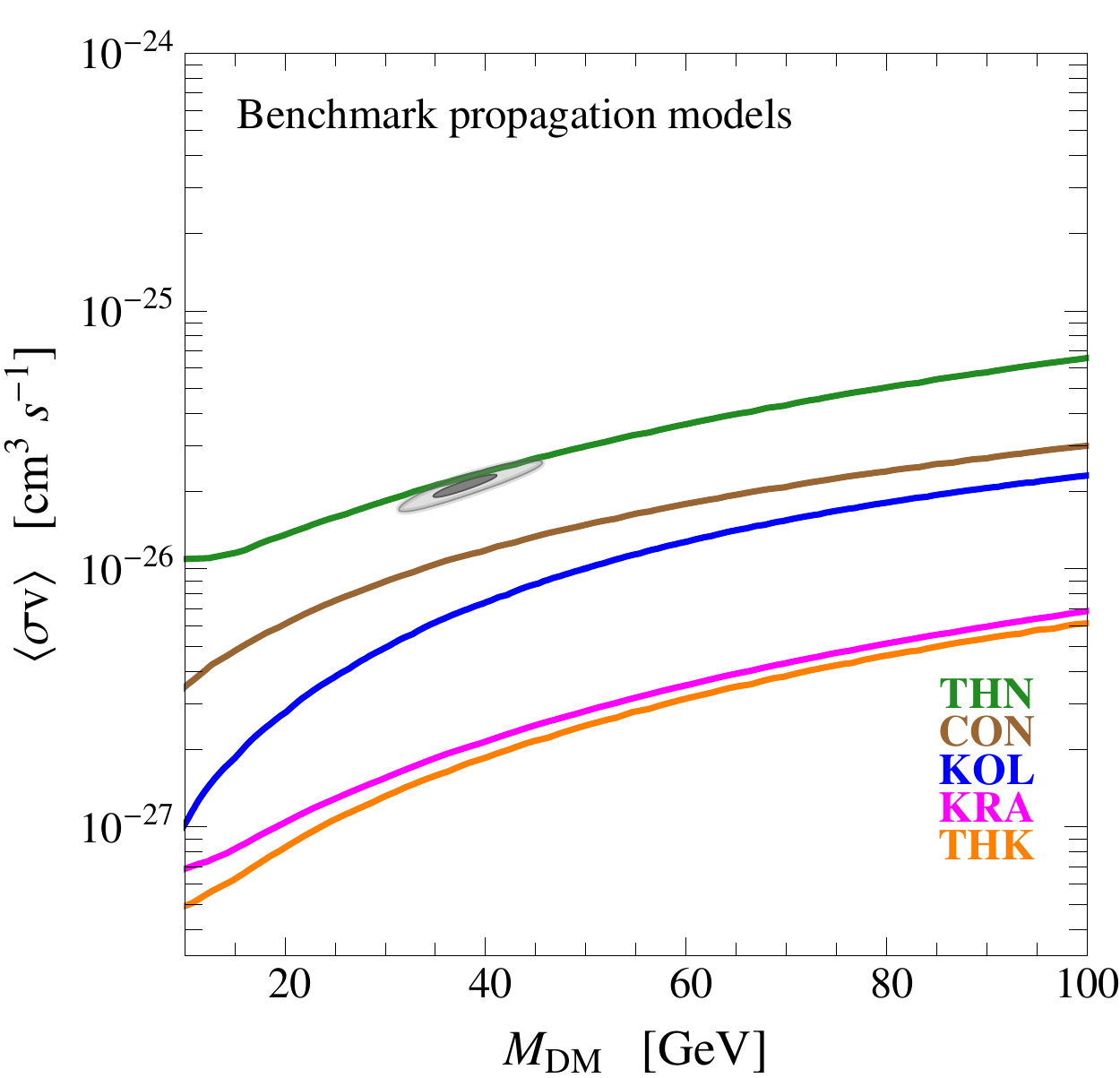}
\endminipage
\minipage{0.32\textwidth}
  \includegraphics[width=1.\linewidth]{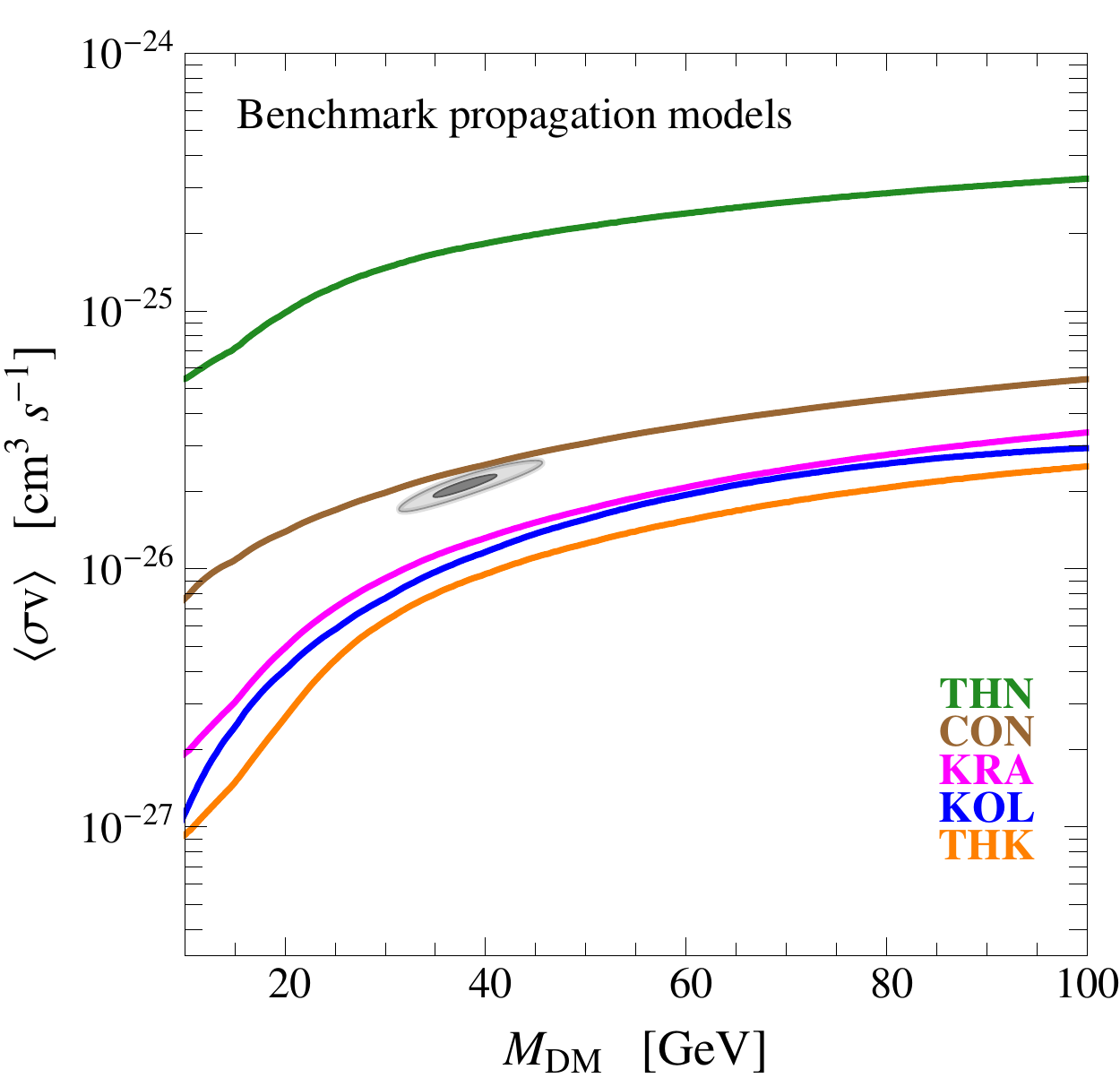}
\endminipage\
\minipage{0.32\textwidth}
  \includegraphics[width=1.\linewidth]{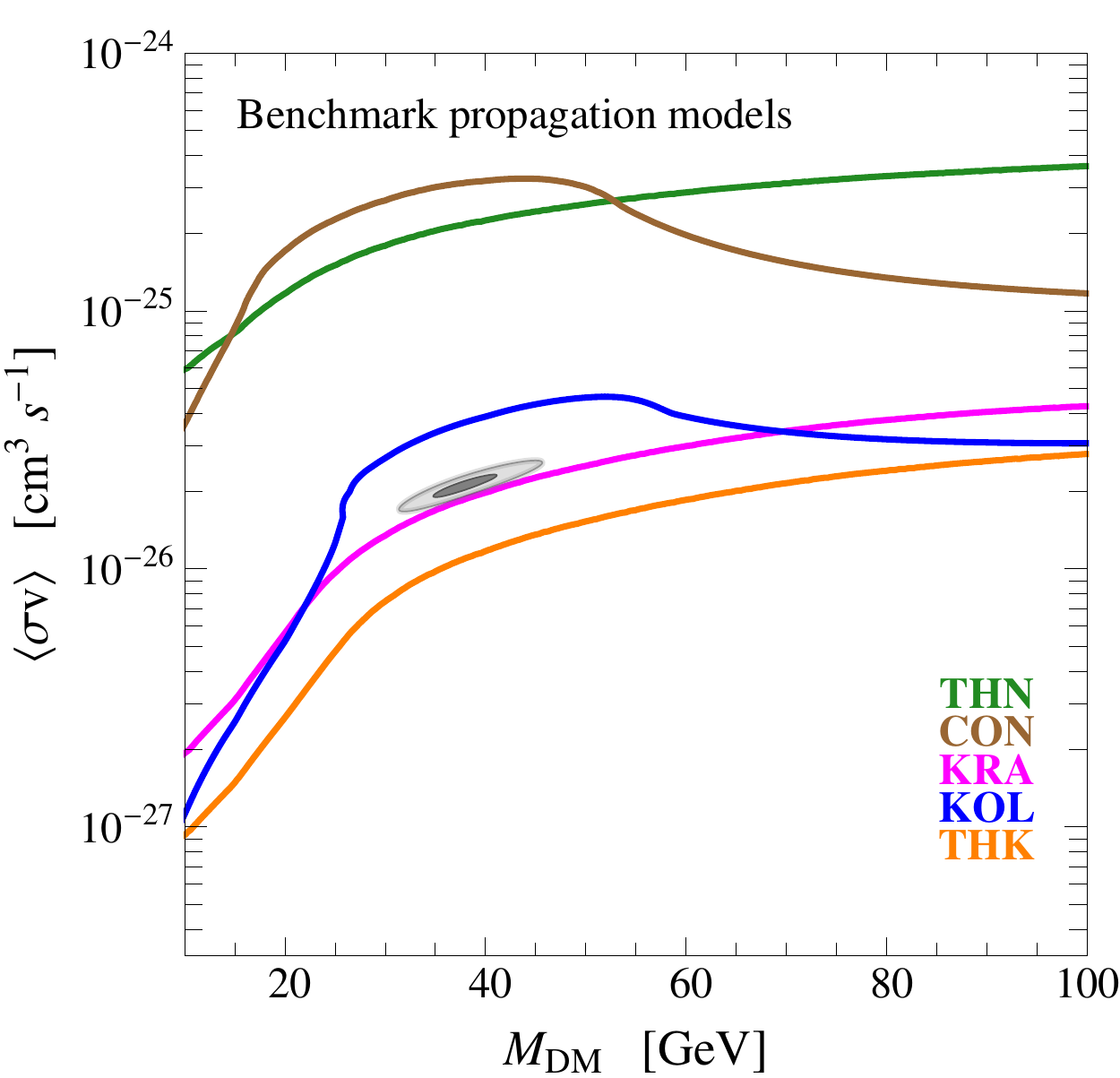}
\endminipage
\caption{\small \textit{ {\bf Left panel}. Antiproton constraints for several propagation models and with the same Fisk potential for protons and antiprotons. {\bf Central panel}. The Fisk potential for the antiprotons $\Phi_{\bar{p}}$ is allowed to vary within 50\% of $\Phi_p$. {\bf Right panel}. $\Phi_{\bar{p}}$ is free.} Plots taken from \cite{cirelli:2014}}
\label{fig:pbar_constraints}
\end{figure}

Nevertheless, in order to set meaningful constraints, the knowledge of CR propagation physics is extremely important.

In particular, I will focus on two aspects:

\begin{itemize}

\item First of all, it is important to know the secondary background with sufficient accuracy. The uncertainties on the propagation parameters discussed in section \ref{sec:prop_parameters} play a major role: In some recent works \cite{evoli:2015,Giesen:2015ufa,Kappl:2015bqa} different estimations of their impact on the antiproton flux is presented (see e.g. Fig.~\ref{fig:pbar_uncertainty}, left panel) and compared to the nuclear physics ones. In this context a precise knowledge of $\delta$ and an accurate determination of the primary proton spectrum is crucial, while the diffusion halo size is not relevant at all.

\item On the other hand, for the (hypothetical) primary component directly coming from DM annihilation, the situation is reversed (see again \cite{evoli:2015} and references therein): Given the different shape of the source function, peaked on the Galactic center, and the large distance the particles have to cross before arriving at Earth, the halo size is, in this case, the major source of uncertainty on the predicted flux. Moreover, especially for low masses, the solar modulation strongly affects the shape of the spectrum. While it is common practice, given a propagation model, to fix the modulation potential $\Phi_{\bar{p}}$ equal to $\Phi_{p}$, the charge-dependent effects -- discussed in section \ref{sec:modulation} -- add a further major source of uncertainty and therefore alter the results.

\end{itemize}

Bearing these considerations in mind, let us focus on the GC excess claim, and see how all these aspects interplay together.

We refer to \cite{cirelli:2014}, and we show in Fig. \ref{fig:pbar_constraints} the constraints for a set of propagation models, already used in \cite{evoli:2012u} and bracketing the parameter range allowed by pre-AMS measurements. The exercise is performed three times, with different choices for the relationship between $\Phi_{\bar{p}}$ and $\Phi_{p}$: {\it 1)} No charge-dependence effect; {\it 2)} Relative difference up to $50\%$, as suggested by a scan performed with the {\tt HelioProp} code over a wide range of Heliospheric propagation parameters; {\it 3)}  $\Phi_{\bar{p}}$ is independent on $\Phi_{p}$.

The reader can easily notice how difficult it is to draw a conclusive statement out of those bounds. 
However, it is worth to point out that the most reliable models, compatible with the large halo size suggested by synchrotron measurements (see Sec. \ref{sec:halo_size}), rule out the DM scenario for most cases; moreover, under the most realistic assumptions on $\Phi_{\bar{p}}$, only the (unlikely) thin halo scenarios keep the DM case alive. 
Nevertheless, the uncertainty due to all these ingredients still do not allow to state a clear exclusion: this is clear from Fig.~\ref{fig:pbar_uncertainty}, right panel where we show the most conservative bounds, based on the thinnest halo compatible with synchrotron observations, and with the minimal model obtained in \cite{evoli:2015} compatible with PAMELA data.

The recently released AMS data on protons, helium, and preliminary data on B/C, although much more constraining on the $\delta$, do not allow to break the degeneracy between different halo sizes, and therefore the bottom line presented above stays unchanged.

It is also worth noticing that several other channels (and DM masses) may provide a good fit of the GC excess \cite{calore:2015,agrawal:2015}, but the most conservative choices of the parameters always permits to evade the constraints, as shown in \cite{evoli:2015}.


\section{Conclusions}

In this contribution I presented an overview of non-trivial aspects in the phenomenology of CR transport, suggesting the idea that the standard lore of CR production, propagation, and modulation in the Heliosphere needs to be significantly revised, given the large amount of new, accurate datasets that have been released in the latest years. In particular, I focused on: The role of the source distribution (in particular, its properties in the inner bulge, and the spiral structure); possible gradients in both the normalization and rigidity scaling of the diffusion coefficient, charge-dependent effects in the solar modulation process.

I discussed how the problem of DM indirect detection is strictly connected to all these issues. Considering the inner Galaxy gamma-ray excess as a case study, I showed that: 
{\it 1) } It is very hard to robustly assess the existence of an anomaly (namely, an excess) with respect to the astrophysical background, since the actual distribution of CR sources, the properties of CR diffusion and advection in different regions of the Galaxy, and the emitting targets (gas, interstellar radiation field) are plagued by large uncertainty.
This is true in particular for the inner Galaxy, and it is possible to show that the inner Galaxy GeV excess can be reabsorbed under very simple and motivated assumptions on the CR source population near the GC. 
{\it 2)} From the point of view of putting constraints on the DM origin of the this excess, it is useful to look at the antiproton channel. I discussed the most important uncertainties affecting the bound; the halo height turns out to be dominant (and in this perspective a better scrutiny of the independent constraints on this observables, coming e.g. from synchrotron data, plays a major role); moreover, it is compelling to take into account charge-dependent effects since they can relax significantly the bound. Even if the fiducial values for all the parameters involved in both CR and heliospheric transport point towards an exclusion of the DM scenario, the uncertainties are still too large to make a final statement.

\section*{Acknowledgements}

I am indebted to all the collaborators who worked with me on all the topics discussed in this highlight, and warmly thank them all: Marco~Cirelli, Giuseppe~Di~Bernardo, Carmelo~Evoli, Ga{\"e}lle~Giesen, Dario~Grasso, Luca~Maccione, Piero~Ullio, Alfredo~Urbano,  Marco~Taoso, and Mauro~Valli.

\end{document}